\documentclass[aps,prb,showpacs,floatfix,twocolumn,superscriptaddress]{revtex4-1}
\usepackage{graphicx}
\usepackage{bm,amsmath}
\usepackage{dcolumn}
\newcommand{\be}{\begin{equation}}
\newcommand{\ee}{\end{equation}}
\newcommand{\bea}{\begin{eqnarray}}
\newcommand{\eea}{\end{eqnarray}}

\begin{document}
\title{ Pair supersolid of the extended Bose-Hubbard model with atom-pair hopping on the triangular Lattice}

\author{Wanzhou Zhang}
\email{zhangwanzhou@tyut.edu.cn}
\affiliation{College of Physics and Optoelectronics, 
Taiyuan University of Technology Shanxi 030024, China}
\author{Yancheng Wang}
\email{wangyancheng@mail.bnu.edu.cn}
\affiliation{ Physics Department, Beijing Normal University, Beijing 100875, China}

\author{Wenan Guo}
\email{waguo@bnu.edu.cn}
\affiliation{ Physics Department, Beijing Normal University, Beijing 100875, China}
\affiliation{ Kavli Institute for Theoretical Physics China, CAS, Beijing 100190, China}

\date{\today}
\begin{abstract}
We  systematically study an extended Bose-Hubbard model with atom hopping and
atom-pair hopping in the presence of a three-body constraint on the 
triangular lattice. 
By means of large-scale Quantum Monte Carlo simulations, the ground-state phase 
diagram are studied. 
We find a continuous transition between the atomic superfluid phase and the 
pair superfluid when the ratio of the atomic hopping and the atom-pair hopping 
is adapted.  We then focus on the interplay among the atom-pair hopping, 
the on-site repulsion and the nearest-neighbor repulsion. 
With on-site repulsion present, we observe first order transitions between the
Mott Insulators and pair superfluid driven by the pair hopping. 
With the nearest-neighbor repulsion turning on, three typical solid phases 
with  $2/3$, $1$ and $4/3$-filling emerge at small atom-pair hopping region.  
A stable pair supersolid phase is found at small on-site repulsion.
This is due to the three-body constraint and the pair hopping, which 
essentially make the model a quasi hardcore boson system.  
Thus the pair supersolid state  emerges basing on the order-by-disorder 
mechanism, by which hardcore bosons avoid classical frustration on the 
triangular lattice. 
The transition between the pair supersolid and the pair superfluid is first 
order, except for the particle-hole symmetric point.
We compare the results with those obtained by means of mean-field analysis.
 
\end{abstract}
\maketitle
\section{Introduction}
\label{sec:intro}
The development in experimentally manipulating ultra-cold atoms in an
optical lattice paved the way to simulate strongly interacting systems in 
condensed-matter physics \cite{qgas}. 
It provides very clean and tunable systems to 
study quantum phase transitions and exotic quantum states, e.g., superfluid, 
Mott insulator, which are not easily accessible in condensed matters. 
The condensation of paired electrons, which provides the basis of 
superconductivity of metallic superconductor, 
plays an essential role in modern condensed-matter physics.
Thus realizing pairing related novel quantum states in the context of 
ultra-cold atoms has attracted considerable recent interest, both in 
theoretical and experimental research. 

One candidate to achieve such states is lattice bosons with attractive on-site 
interactions, which is stabilized by a three-body constraint
\cite{zoller1, th1}. The three-body constraint has been realized
by large three-body loss processes\cite{3bodyloss, 3body2}. The system can be 
mapped into spin-one atoms at unit filling\cite{sp1mott}.
Besides the conventional single atom superfluid (ASF) state,
a pair (dimer) superfluid (PSF) phase consisting of the 
condensation of boson pairs emerges under sufficiently strong 
attraction\cite{zoller1,th1}. 
The PSF state is manifested as a second-order effect in the atom hopping in
the optical lattice.  Various phase transitions among  
the atomic superfluid (ASF), Mott insulator (MI) and PSF are investigated
in great detail \cite{zoller1, th1, mfy1,mfy2,psf,ycc,fintemp}. Both the 
ground state and the thermal phase diagrams are obtained. 
Although the pair supersolid (PSS) state was predicted in system with 
correlated hopping \cite{Schmidt, hcj} and 
the paired two-species bosons supersolid was predicted on the square 
lattice\cite{pch} with the attractive interaction between different species  
and the repulsion between the same species of atoms turning on,
the pair (single species) supersolid was not found in the present system 
when the nearest-neighbor repulsion is included, 
except for an isolated continuous supersolid at the Dirac point \cite{zoller1}.
The reason might be the same instability of the
supersolid state (SS) on the square lattice, on which the former research 
focused, for hardcore bosons\cite{ssm3, sshcsq}.  

Another practical way to access pairing phenomena is considering repulsive 
bound-atom pairs in optical lattice, which have been
realized in experiments \cite{ex1,pair}. 
The system can be reasonably described by explicitly including atom-pair 
hopping\cite{liang2,liang3} in the ordinary extended Bose-Hubbard model (EBH).
Such a pair hopping can also be introduced in atom-molecule
coupling system on the state-dependent optical lattice
\cite{xfzhou}, or by a mechanism based on transport-inducing 
collisions\cite{cor1}.  The resulted Bose-Hubbard Hamiltonian is thus 
\begin{eqnarray}
H &=&-\sum\limits_{i}\mu n_{i}+\frac{U}{2}\sum\limits_{i}n_{i}(n_{i}-1) 
+\sum\limits_{\langle i,j \rangle} V n_{i}n_{j} \nonumber\\
&& -t \sum\limits_{\langle i,j \rangle} (a_{i}^{\dag }a_{j}+a_{i}a_{j}^{\dag})
-t_p\sum\limits_{\langle i,j\rangle} (a_{i}^{\dag 2} a_{j}^2+a_{i}^2 a_{j}^{\dag 2}), 
\label{eq1}
\end{eqnarray}
where $a^{\dag}_i$ ($a_i$) creates (annihilates) a boson at site $i$,
$t$ is the atom hopping amplitude for nearest neighbor sites $\langle ij \rangle$,  
$t_p$ is  the pair-hopping amplitude, $V$ the nearest-neighbor repulsion, 
$\mu$ the chemical potential, $U>0$ the on-site repulsion. The three-body 
constraint is applied which requires the occupation number $n_i=0, 1,$ or 2.
It is naturally expected that 
PSF, MI and solid phases should emerge at certain parameters. 

The three-body constraint and the pair hopping makes the system resemble 
hardcore bosons.  For hardcore bosons supersolid state  
emerges on the triangular lattice basing on an order-by-disorder mechanism, by 
which a quantum system avoids classical frustration\cite{tri1,tri2}.
Aiming to realize PSS state, we thus focus on the triangular lattice.
A preliminary mean-field (MF) analysis \cite{wzg} does predict the PSS phase
in such a system.
In present work we systematically study the model by means of a large scale 
Quantum Monte Carlo simulation. 
The phase diagram of the system  is studied in great detail.  
With the nearest-neighbor interaction turning on,  we report three types of 
solid phases at small hopping strength.  Increasing pair hopping $t_p$, we 
find the expected PSS phase. 

This paper is organized as follows: We first discuss the ground state at the 
classical limit in Sec.\ref{sec:class}, which is useful to identify 
various solids and MI states, in which we are interested. 
We describe the quantum Monte Carlo (QMC) method and useful observables 
in Sec. \ref{sec:method}.  
We then present, in Sec.\ref{sec:results}, QMC simulation results, 
comparing with the mean-field analysis.  The results for the noninteracting 
($U=V=0$) case are described  in Sec.  \ref{sec:u0v0}.  
Those for the interacting case $U\ne 0$, $V = 0$  
are given in Sec. \ref{sec:une0v0} and for the case $V \ne 0$ 
in Sec. \ref{sec:vne0}, 
focusing on the parameter regions where the pair supersolid phase emerges. 
The the nature of related phase transitions are discussed. 
We conclude in Sec. \ref{sec:discon}  with discussions.

\section{Classical Limit} \label{sec:class}

\begin{figure}[htpb]
\includegraphics[width=6cm, height=8cm]{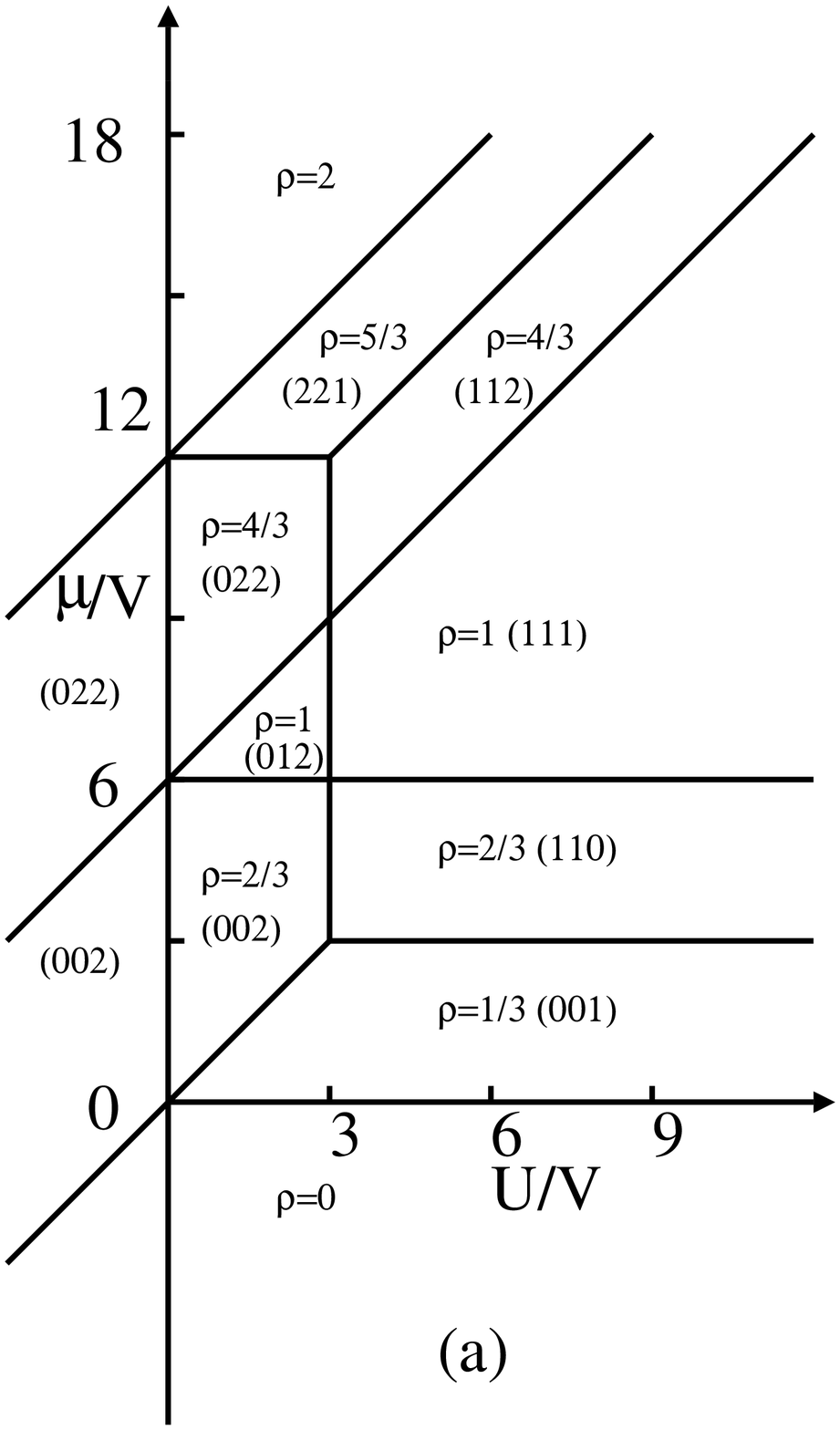}
\vskip 0.2cm
\includegraphics[width=2.5cm]{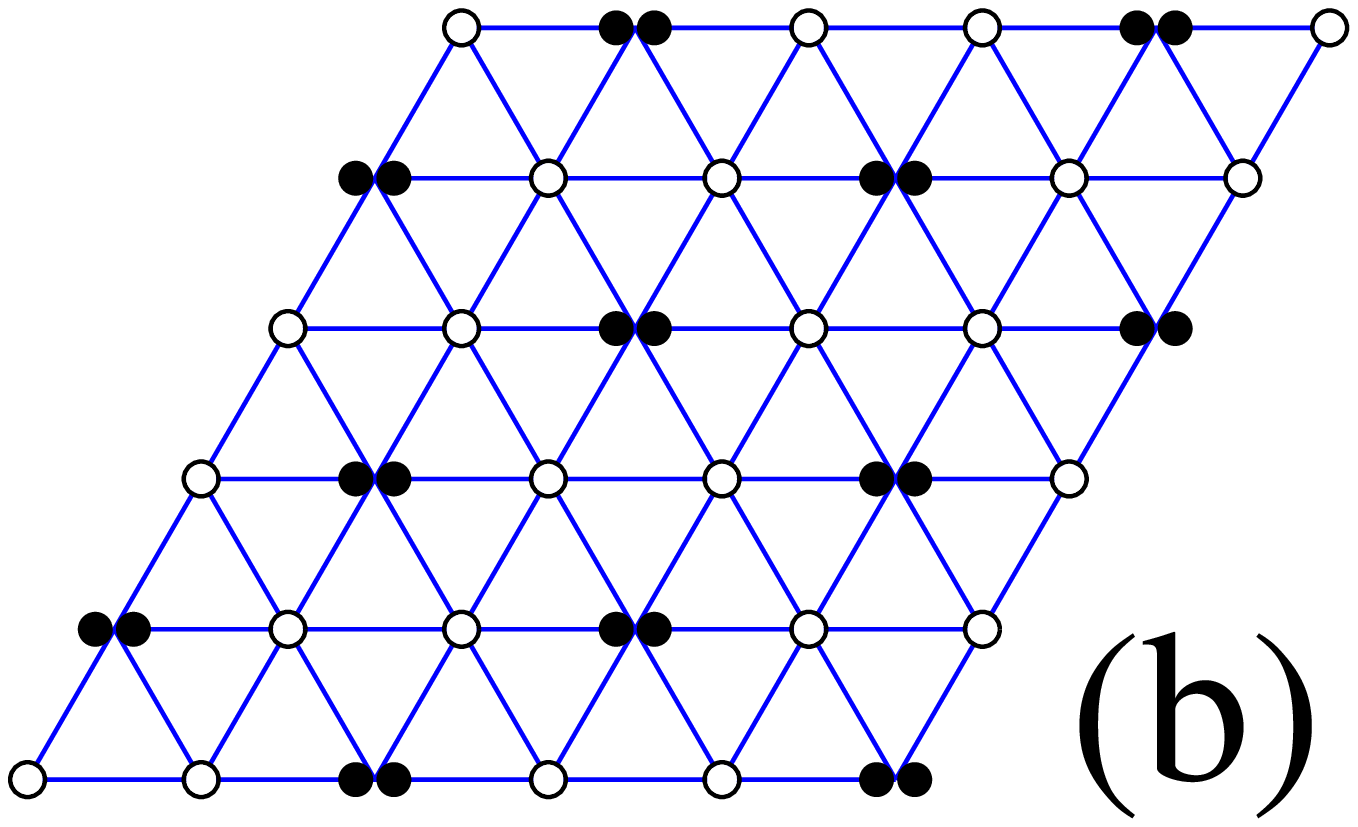}
\includegraphics[width=2.5cm]{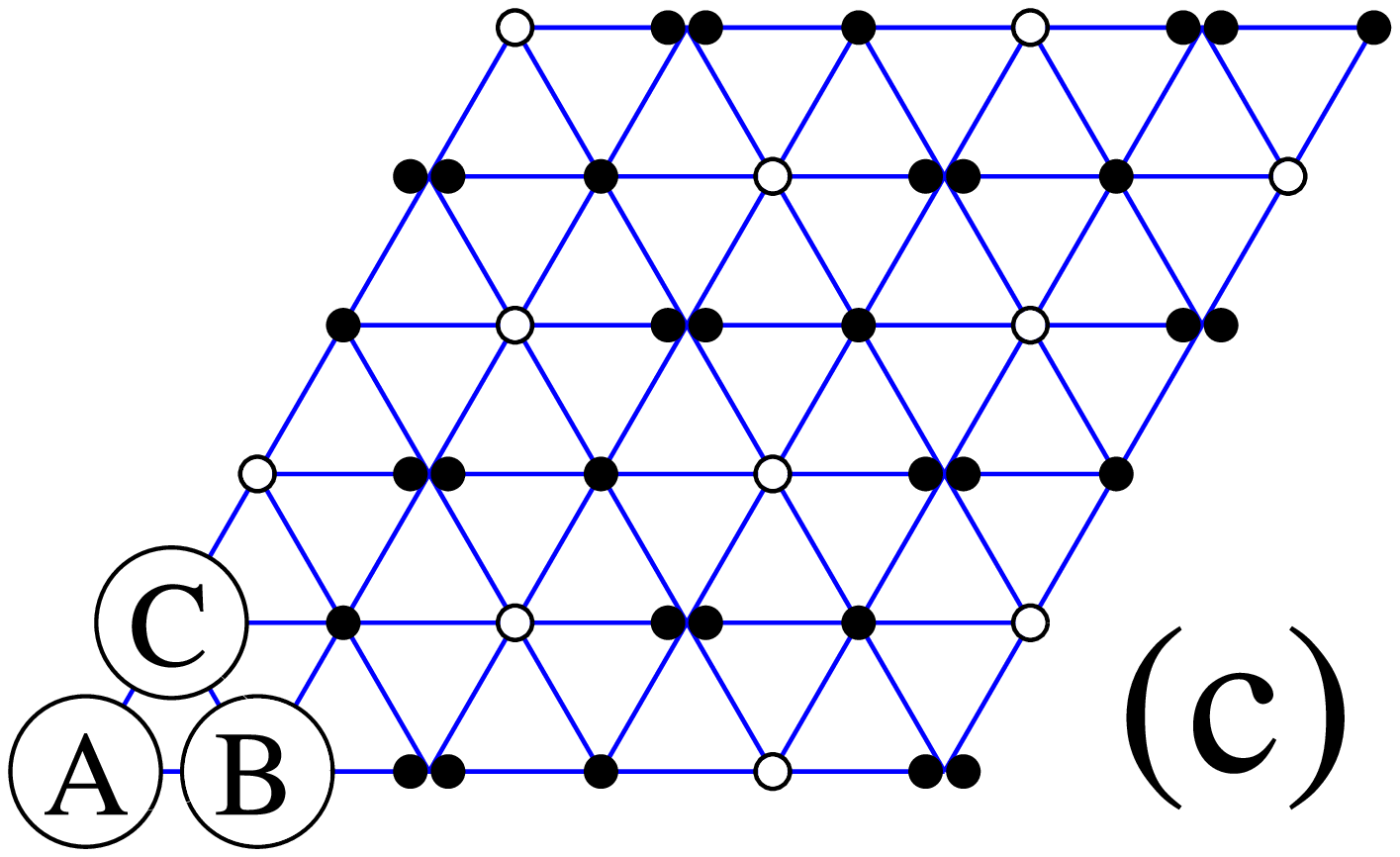}
\includegraphics[width=2.5cm]{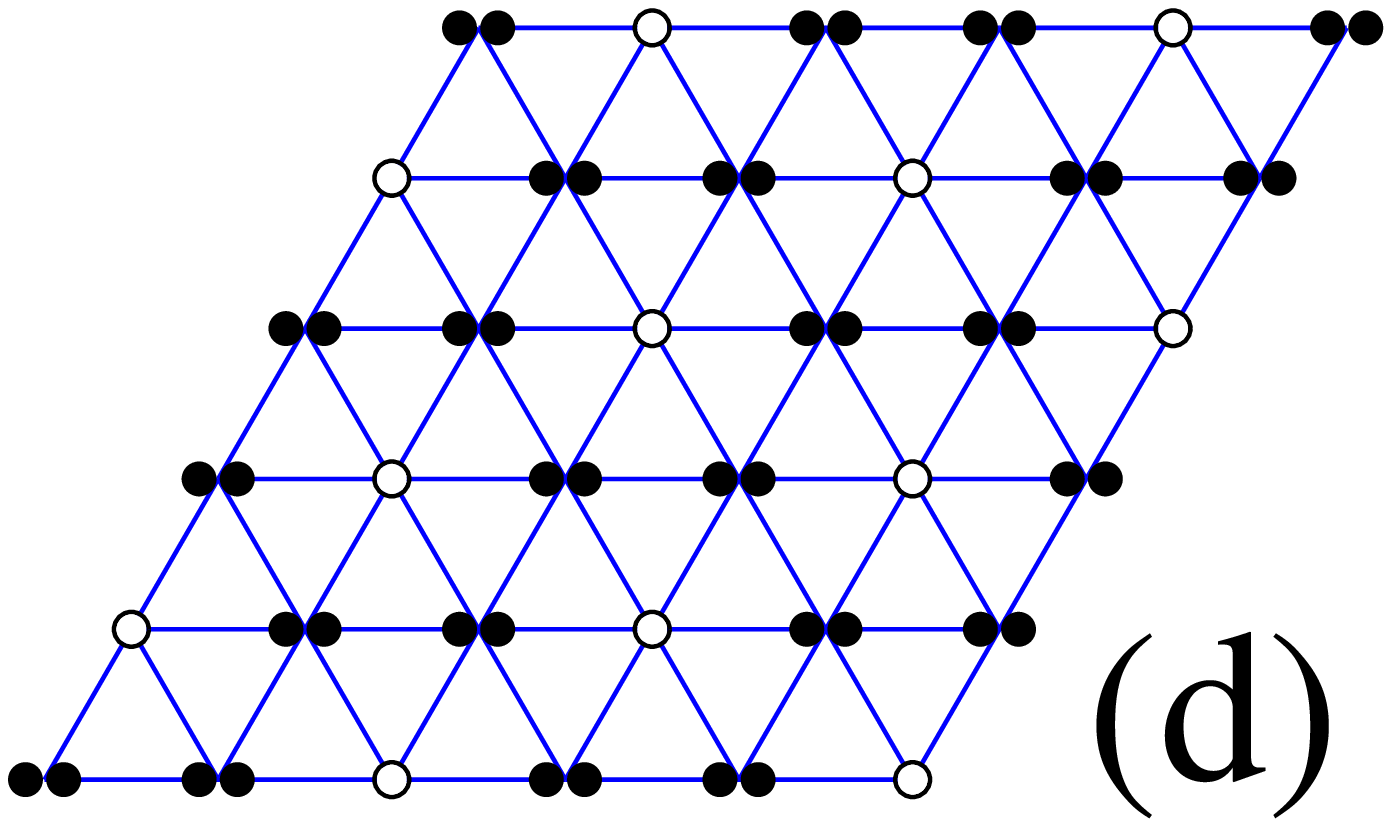}
\caption{(Color online) (a) The ground-state ($U/V-\mu/V$)  phase diagram in 
the classical  limit $t=0, t_p=0$.
$(n_A, n_B, n_C)$  denotes a solid order in a $\sqrt{3} \times \sqrt{3}$ 
ordering with wave vector ${\bf Q}=(4\pi/3,0)$, 
where the numbers are boson occupations on three sublattices, respectively. 
(b) SI phase (0, 0, 2) with filling $\rho=2/3$.
(c) SII phase (0, 1, 2) with filling $\rho=1$.
(d) SIII phase (0, 2, 2) with filling $\rho=4/3$. }
\label{clph}
\end{figure}

We first study the zero temperature phase diagram in the classical limit ($t=0, t_p=0$)
in the presence of the three-body constraint.

At half filling, the solid order is frustrated on the triangular lattice, and the classical
model has a hugely degenerate ground state with an extensive zero-temperature entropy \cite{Wannier}.
We focus on the solid ordering, away from half filling, with wave vector ${\bf Q}=(4 \pi/3,0)$. 
The lattice is divided into three sublattices: $A, B, C$. The solid order is thus denoted as $(n_A, n_B, n_C)$ 
or its equivalent permutations,
where the numbers are the boson occupations on three sublattices, respectively. 
The energy per site is 
\bea
e&=&-\mu \frac{n_A+n_B+n_C}{3} + V (n_A n_B+n_A n_C+n_B n_C) \nonumber \\
 &+&U \frac{n_A(n_A-1)+ n_B(n_B-1)+ n_C(n_C-1)}{6} . 
\eea
By comparing energy per site,
we obtain phase boundaries between 
various phases, as shown in Fig. \ref{clph}.  Three atom-pair solid phases 
appearing in the region $0<U/V<3$ are of special interest, which have 
two bosons sitting on one or two sublattices, as shown in Fig.\ref{clph}(b), 
(c) and (d). They are denoted as SI, SII and SIII, with densities 
$\rho=2/3, 1, 4/3$, respectively.
Other solid states related with configurations (0, 0, 1), (1, 1, 0), (1, 1, 2),
 (1, 2, 2) and MI states (1, 1, 1) and (2, 2, 2) are also found.

\section{method and observables}
\label{sec:method}
We simulate the model (\ref{eq1}) using the stochastic series expansion (SSE)  
QMC method\cite{sse} with the directed loop update\cite{direcloop}, which is 
improved for the pair hopping.
The head of a directed loop  carries a create (annihilate) operator 
$a (a^{\dag})$ in the conventional directed loop algorithm for the BH model 
with  single particle hopping. 
We improve the algorithm by allowing the head of a directed loop carry a pair
create (annihilate) operators  $a^{2\dag}$($a^{2}$) for the present EBH model 
with additional pair hopping terms.  Similar improvement has been described in 
the literature\cite{two-loop}.
In the simulations, temperature is chosen as $\beta=L$ to ensure the system 
sitting in its ground state. 

To distinguish the ASF and the PSF states,
we  define an even (odd) superfluid stiffness  $\rho_s^{(\alpha)}$ as the order 
parameter \cite{sfs}:
\begin{equation}
\rho_{s}^{(\alpha)}=\frac{\langle W(\alpha)^2\rangle}{4\beta(4t_p+t)},
\end{equation}
where $\beta$ is the inverse temperature, $\alpha$ can be `even' or  `odd'; 
$W(\alpha)$ is the total even (odd) winding number. 
The factor  `$4$' multiplying  $t_p$ is due to the hopping of paired atoms 
\cite{sand}. The total superfluid stiffness is $\rho_s \equiv \rho_s^{(even)}+
\rho_s^{(odd)}$.
For a PSF state,  we define the pair superfluid order parameter 
$\rho_s^{(p)} \equiv \rho_{s}^{(even)}-\rho_s^{(odd)}>0$, 
while for an ASF state, $\rho_{s}^{(p)}=0$, but $\rho_s \ne 0$.  
It is worthy to note that, without single atom hopping, $\rho_s^{(odd)}$ is 
guaranteed to be zero.  

The structure factor is defined to characterize the solid  order:
\begin{equation}
S({\mathbf Q})/N=\langle \rho_{{\mathbf Q}} \rho^{\dagger}_{{\mathbf Q}} \rangle,
\end{equation}
where
$\rho_{{\mathbf Q}}=(1/N)\sum_i n_i \exp(i {\mathbf Q} {\mathbf r_i})$.
The solid states SI, SII and SIII share the same ordering at
the wave vector ${\mathbf Q}=(4\pi/3, 0)$, but  bear  
different values  $4/9$, $1/3$, $4/9$, respectively, in the 
perfect ordering. 

\section{results}
\label{sec:results}
\subsection{
Non-interaction case: $U=0$,$V=0$;
Competition between ASF and PSF 
}
\label{sec:u0v0}
\begin{figure}[htpb]
\vskip 0.2 cm
\begin{center}
 \includegraphics[width=8.5cm ]{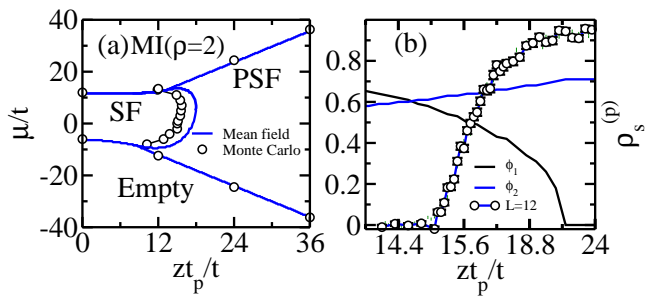}
\end{center}
\caption{(Color online)  
 (a) ($z t_p/t, \mu/t$) phase diagram of model (\ref{eq1}) obtained by means of
the MF (solid blue lines) and the QMC(circles) method, which contains Empty, 
MI, SF, and PSF phases at $U=0$, $V=0$. 
(b) Pair superfluid order parameter $\rho_s^{(p)}$ as a function of 
$z t_p/t$ for $\mu=0$ ($L=12$).
The MF order parameters $\phi_1$ and $\phi_2$ are also shown.  
}
\label{ph0}
\end{figure}

The ASF-PSF  transition  for model (\ref{eq1})  without on-site repulsion $U$ 
and nearest-neighbor repulsion $V$ was 
previously discussed by means of MF analysis\cite{wzg}. 
Here we present our QMC results for such a system. To demonstrate the 
differences between our results and the MF results, we show both of them in 
Fig. \ref{ph0}. 
We find an Empty phase and a $\rho=2$ MI phase at 
negative and large chemical potentials, respectively. The presence of the 
MI state is due to the three-body constraint. Between the Empty phase 
and the MI phase,  there are two SF phases: an ASF phase and a PSF phase. 

With small pair hopping $t_p \approx 0$, the system  exhibits an Empty-ASF 
transition  at $\mu=-6t$ and a SF-MI transition at $\mu=12t$.  
This can be understood in the single particle picture:
to put a boson on the empty lattice gains potential energy $-\mu$ and kinetic 
energy $-zt$, where $z=6$ is the coordination number of the triangular lattice.
Adding a hole to the MI ($\rho=2$) state costs chemical energy $\mu$, while 
obtain kinetic energy $-2zt$. 
At large pair hopping $t_p \gg t$, the system shows a PSF-Empty transition 
at $\mu/t=-6 t_p/t$.  A pair of bosons emerging on the Empty phase gets 
$-2zt_p$ kinetic energy and potential energy $-2\mu$.  
Similarly, the PSF-MI transition line is at $\mu/t=6 t_p/t$ by analyzing the 
emerging of a pair of holes on the MI state.

To demonstrate the transition between the ASF and the PSF states in more 
detail, we show here the system behaviors along $\mu=0$ in Fig. \ref{ph0}(b).
The PSF state was predicted at $t_p/t>2.9$ in the MF frame using the criterion 
$\phi_1 \equiv \langle a \rangle = 0$ but $\phi_2 \equiv \langle a^2 \rangle 
\ne 0$ \cite{wzg}.  
This transition is confirmed by our unbiased QMC simulations. However, the 
transition point is $t_p/t=2.5$. 
The PSF state is characterized by $ \rho_{s}^{(p)} >0$.
The phase transition between ASF and PSF is continuous. 
In the presence of strong on-site attractive interactions ($U<0$), the PSF has 
been predicted in the lattice bosons with a three-body hard-core 
constraint\cite{psf, mfy2}.  However, the transition between ASF 
and PSF phases is claimed to be first order \cite{mfy2}.

\subsection{MI-PSF phase transitions for $U > 0$ and $V=0$}
\label{sec:une0v0}
\begin{figure}[htpb]
\vskip 0.8cm
\includegraphics[width=8.5cm]{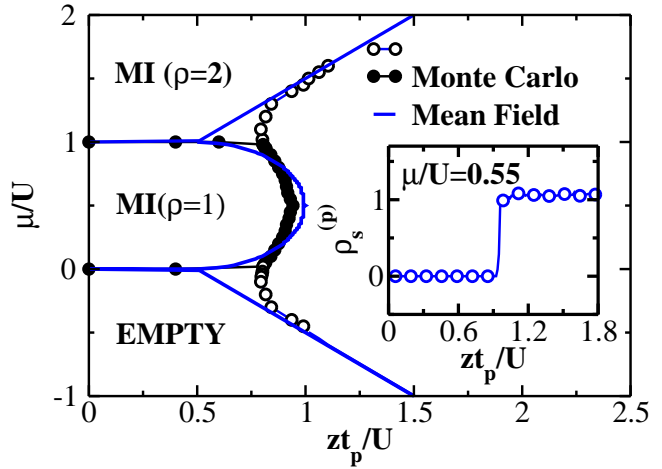}
\caption{(Color online) $(zt_p/U,\mu/U)$ phase diagram obtained by QMC 
(circles) and MF analysis (solid blue lines) 
at $t=0, V=0$, which contains the Empty, MI ($\rho=1$), MI ($\rho=2$), and PSF 
phases at $\mu/V=0.55$. 
Open and solid circles correspond to continuous and first order transitions,
respectively. 
The pair superfluid order parameter $\rho_s^{(p)}$ vs. $z t_p/U$ is
shown in the inset. 
}
 
\label{une0v0ph}
\end{figure}
In the presence of the on-site repulsion $U$ and the absence of the single 
atom hopping $t$, the ASF state disappears, instead an MI($\rho=1$) phase 
emerges.  Without three-body constraint, the general picture of the phase 
diagram has been studied by Zhou {\it et al} using the MF method
\cite{xfzhou}.
We show here the phase diagram, under three-body constraint, obtained by using 
the MF method \cite{wzg} and by the QMC simulations in 
Fig. \ref{une0v0ph}.  
The phase boundaries obtained by using the two methods are in good agreement 
for large $t_p$. The Empty-PSF phase transition  line is straight with the 
slope  $\mu/(zt_p)=-1$, which  
can be understood in the single particle picture: A boson pair emerging in
the empty phase gets $-2 zt_p $ kinetics energy and $-2\mu$ potential energy. 
Similarly, the MI($\rho=2$)-PSF boundary is also straight with a slope $\mu/(zt_p)= 1$ 
at large $t_p$.
QMC simulations provide more accurate boundaries between the   
MI phases and PSF phase.  The $\rho=2$ MI-PSF transition 
is found to be continuous and the MI($\rho=1$)-PSF transition is first order, 
as illustrated in the inset of Fig. \ref{une0v0ph}.  

\subsection{ Pair supersolid phase for $U=0$ and $U>0$}
\label{sec:vne0}

\begin{figure}[htpb]
\includegraphics[width=8cm]{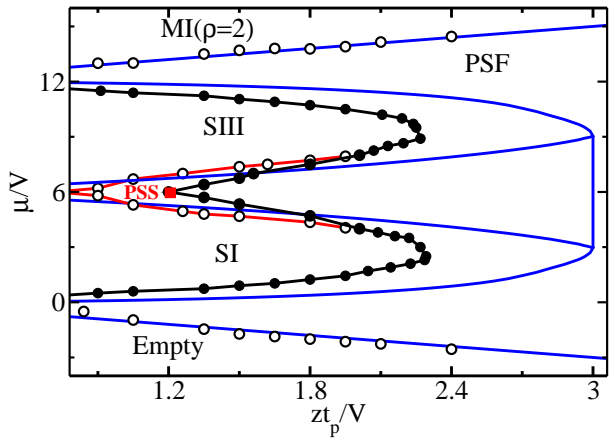}
\caption{(Color online)  $( zt_{p}/V, \mu/V )$ phase diagram obtained by QMC 
(circles) and the MF analysis (solid blue lines)  of model (\ref{eq1})  
at $t=0, U=0$, which contains the Empty, MI, SI, SIII, PSF and PSS phases.
Open and solid circles correspond to continuous and first order transitions,
respectively. 
The PSS-PSF transition is first order except for the $\mu/V=6$ point
marked by a red square \cite{1st}.
} 
\label{u0v1he}
\end{figure}
As a result of the absence of on-site repulsion, boson pairs easily
emerge on a site without costing of any potential energy. 
In the limit  $t_p/t \to \infty$ (or $t=0$), boson pairs (hole pairs) can 
easily hop on the lattice, forming a PSF phase.
When the nearest-neighbor interaction $V$ present, solid ordering appears naturally.
Hardcore bosons with single-atom hopping on the triangular lattice has an SS phase between two 
solid phases ((0, 0, 1) and (0, 1, 1)) \cite{tri1,tri2}.
Bosons on the triangular lattice with pair hopping under three-body constraint 
mimic such behavior with the two solid  phases replaced by the (0, 0, 2) and 
(0, 2, 2) solid phases respectively.
The global phase diagram obtained by means of the MF method \cite{wzg} and QMC 
simulations is shown  in Fig \ref{u0v1he}. 
Both method predict a PSS phase between the two solid phases. However, the PSS
region is much smaller in the QMC phase diagram. This is due to 
quantum fluctuations ignored in the MF analysis. 

We further illustrate the phase diagram along two lines: one is 
$z t_p/V=1.05$, the other is $\mu/V=7.5$.

\begin{figure}[htpb]
\includegraphics[width=8cm]{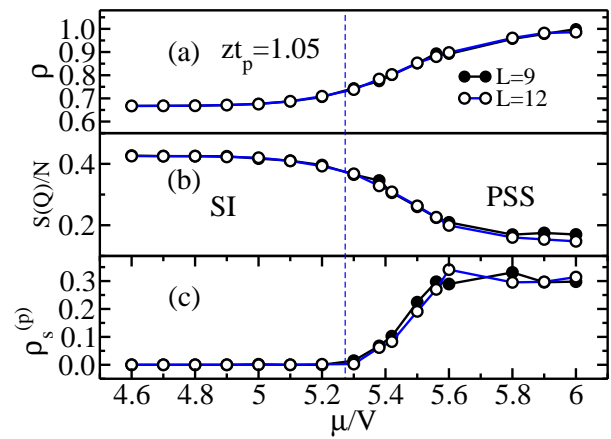}
\caption{(Color online) 
The density $\rho$, structural factor $S({\bf Q})/N$, PSF stiffness 
$\rho_s^{(p)}$ as functions of  $\mu/V$  at  $z t_p=1.05$.}
\label{ztp1.05}
\end{figure}

Figure \ref{ztp1.05} shows the density $\rho$, structure factor $S({\bf Q})/N$ 
and pair superfluid order parameter $\rho_s^{(p)}$ as functions of $\mu/V$ 
along the $z t_p/V=1.05$ line.
Considering the hole-particle symmetry, we only present results for $\mu/V \le 6$.
The PSS state emerges clearly between the two solid phases (SI and SIII), which
 is further proved by a finite-size scaling analysis of the structural factor 
and the pair superfluid stiffness $\rho_s^{(p)}$ (not shown). 
\vskip 1cm

\begin{figure}[htpb]
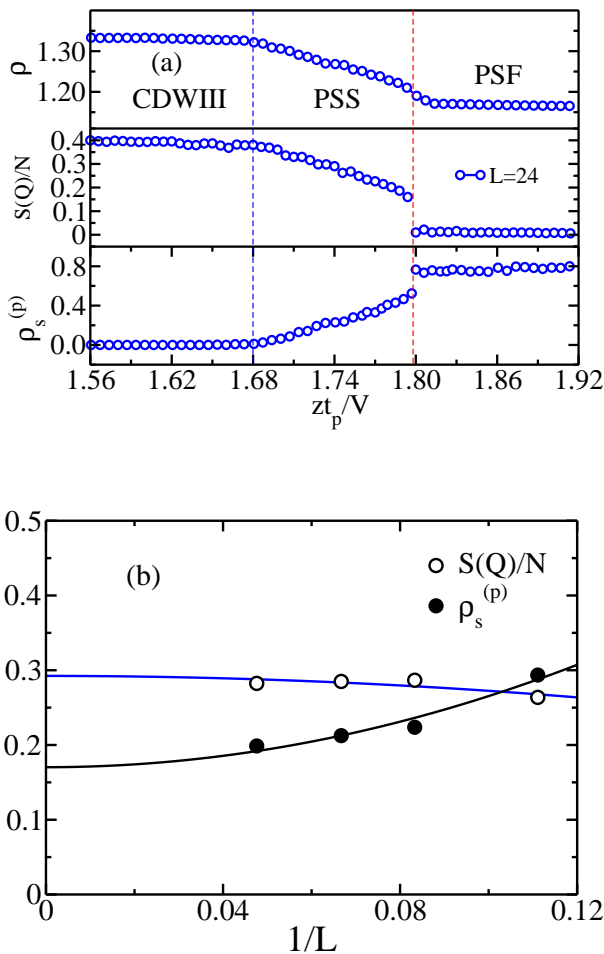

\includegraphics[width=8cm]{fig5-1.eps}
\vskip 1.0cm
\includegraphics[width=8cm]{fig5-2.eps}
\caption{(Color online) (a) The density $\rho$, structural factor 
$S({\bf Q})/N$, and  PSF 
order parameter $\rho_s^{(p)}$ as functions of  pair hopping 
$z t_p/V$ at $\mu/V=7.5$. 
(b) Finite size scalings of the structural factor $S({\bf Q})/N $ and PSF 
order parameter $\rho_s^{(p)}$ at $z t_p/V=1.74$ and  $\mu/V=7.5$, 
showing PSS character at thermodynamic limit. } 
\label{t_p0.29mu7.5}
\end{figure}

Next, we scan the pair hopping $t_p$ along $\mu/V=7.5$.
The density $\rho$, structural factor $S({\bf Q})/N$ and pair superfluid 
order parameter $\rho_s^{(p)}$ as functions of  the pair hopping $t_p$ are 
shown in Fig. \ref{t_p0.29mu7.5}(a), with the PSS state emerging in the region 
$1.668<z t_p/V<1.788$. 
Figure \ref{t_p0.29mu7.5}(b) shows a finite size scaling analysis 
of the pair superfluid order parameter $\rho_s^{(p)}$ and the structural 
factor $S({\bf Q})/N$ at $z t_p/V = 1.74$, $\mu/V=7.5$,
which proves that the state is indeed a pair supersolid.

The phase transition between the PSS and other states are interesting. 
The PSS-SI(SIII) transition is continuous from  continuous variation of the 
density  and the structure factor, while the PSS-PSF transition is first 
order, except for the particle-hole symmetric point \cite{1st}.
as illustrated by the jump of the structural factor and the PSF order parameter
in Fig. \ref{t_p0.29mu7.5}(a). 
From  the jumps of the PSF order parameter and the structural factor 
(not shown), the PSF-solid phase transitions are also first order. 

\begin{figure}[htpb]
\vskip 1cm 
\includegraphics[width=8.cm]{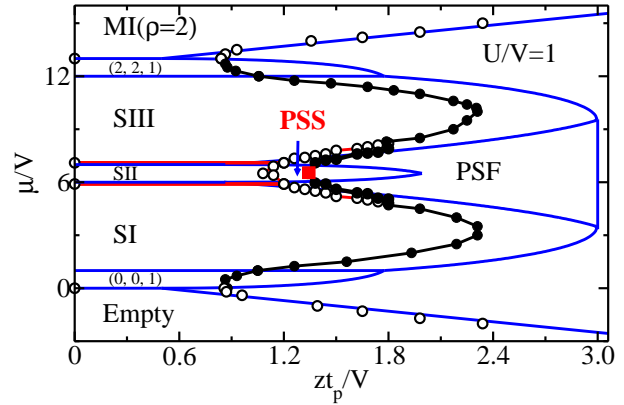}
\caption{ (Color online)  $(zt_p/V, \mu/V)$ phase diagram obtained by QMC 
(circles) and the MF (solid blue lines) method at $U/V=1$, which contains  
the Empty, MI($\rho=1, \rho=2$), SI, SII, SIII, (0, 0, 1) , (2, 2, 1), PSF 
and PSS phases.
Open and solid circles correspond to continuous and first order transitions,
respectively. The red square marks the special high symmetric point.
}
\label{u1v1he}
\end{figure}

We now check whether the pair supersolid phase could survive as a 
weak on-site repulsion turns on. 
Figure \ref{u1v1he} shows  the MF and QMC phase diagram 
at weak repulsion $U/V=1$.
Three new solid phases, i.e., SII, (0, 0, 1) and (2, 2, 1),  emerge comparing 
to the case $U=0$.
The PSS phase persists. 
Increasing $t_p$, all solid phases melt into the PSF phase 
eventually. 
The persistence of the PSS state at weak on-site repulsion is because that 
the PSS state depends strongly on the frustrated 
geometrical structure of the triangular lattice for hardcore bosons:
although the maximum  occupation is two, the pair hopping essentially make 
the model (\ref{eq1}) a quasi 
hardcore boson system. Weak on-site repulsions do not change this feature. 

At the large  $U/V$ limit, the phase diagram should reduce to the one 
shown in Fig. \ref{une0v0ph}. This suggests the PSS phase will disappear
finally. Indeed, we don't find PSS phase when $U/V \ge 4$.

\section{Conclusion}
\label{sec:discon} 

We have studied the ground-state phase diagram of the 
extended Bose-Hubbard model with explicit atom-pair hopping terms, under
the three-body constraint, on the triangular lattice. 
Experimentally, the atom-pair hopping physics has been realized.
The triangular optical lattice can also be implemented by
three beams of lasers\cite{becker}. 
The atom-molecule coupling system \cite{xfzhou} and the spin-1 boson system 
\cite{sp1mott} in state dependent optical triangular lattice 
are also candidates to simulate pair related quantum phenomena.
By means of the improved  SSE QMC method,  we obtain accurate phase diagrams 
at various parameter values.
Rich solid phases and the pair superfluid phase are found, 
as well as the pair supersolid phase,  which
emerges when the nearest-neighbor repulsion and the pair hopping are present, 
under the condition that the on-site repulsion is not large. 
The three-body constraint and 
the pair hopping essentially make the model (\ref{eq1}) a quasi 
hardcore boson system. 
Weak on-site repulsions do not change this feature. 
Therefore, the mechanism  of forming the PSS state is the same as that of 
forming SS state for hardcore bosons on the triangular lattice. 
The properties of phase transitions involved, e.g., 
the ASF-PSF, the MI-PSF, the PSS-PSF, the PSF-solid transitions, are studied. 

It would be interesting to study the finite temperature properties of 
the present model. 
The three-body constraint due to the three-body loss\cite{3bodyloss},
can be suppressed, or even inhibited by the quantum Zeno effect\cite{zeno}.
Thus the quantum Zeno effect provides us a 
basis to study the pair physics without three-body constraint.

\acknowledgments
W.Z. thanks the Found from Taiyuan University of Technology,
the NSF of Shanxi Province, and discussions with Y.-B. Zhang and R.-X. Yin. 
The work is supported by the NSFC under Grant No.11175018.

\end{document}